\documentclass[twocolumn]{aastex701}

\newcommand{\hi}{\ion{H}{1}}

\begin{document}

\title{KDG~162: A Nearby Isolated Star-forming Dwarf Galaxy}

\author[0000-0002-5259-2314]{Gagandeep S. Anand}
\affiliation{Space Telescope Science Institute, 3700 San Martin Drive, Baltimore, MD 21218, USA}
\email{ganand@stsci.edu}  

\author[0000-0003-0307-4344]{Igor D. Karachentsev}
\affiliation{Special Astrophysical Observatory, Russian Academy of Sciences, Nizhniy Arkhyz, Karachai-Cherkessia 369167, Russia}
\email{idkarach@gmail.com}

\author[0000-0001-7531-9815]{Meredith J. Durbin}
\affil{Department of Astronomy, University of Washington, Box 351580, Seattle, WA 98195-1580, USA}
\affil{Department of Astronomy, University of California Berkeley, Berkeley, CA 94720, USA}
\email{mdurbin@uw.edu}

\author[0000-0002-9291-1981]{R. Brent Tully}
\affil{Institute for Astronomy, University of Hawaii, 2680 Woodlawn Drive, Honolulu, HI 96822, USA}
\email{tully@ifa.hawaii.edu}

\author[0000-0002-1691-8217]{Rachael Beaton}
\affiliation{Space Telescope Science Institute, 3700 San Martin Drive, Baltimore, MD 21218, USA}
\affiliation{Department of Physics and Astronomy, Johns Hopkins University, Baltimore, MD 21218, USA}
\email{rbeaton@stsci.edu}

\author[0000-0001-8416-4093]{Andrew Dolphin}
\affiliation{Raytheon, 1151 E. Hermans Road, Tucson, AZ 85756}
\affiliation{University of Arizona, Steward Observatory, 933 North Cherry Avenue, Tucson, AZ 85721, USA}
\email{adolphin@rtx.com}

\author[0000-0001-6927-5473]{Serafim S. Kaisin}
\affiliation{Special Astrophysical Observatory, Russian Academy of Sciences, Nizhniy Arkhyz, Karachai-Cherkessia 369167, Russia}
\email{skai@sao.ru}

\author[0000-0003-0736-7609]{Lidia N. Makarova}
\affiliation{Special Astrophysical Observatory, Russian Academy of Sciences, Nizhniy Arkhyz, Karachai-Cherkessia 369167, Russia}
\email{lidia.makarova@gmail.com}

\author[0000-0002-3234-8699]{Edward J. Shaya}
\affil{Department of Astronomy, University of Maryland, College Park, MD 20743 USA}
\email{eshaya2@gmail.com}

\begin{abstract}

The nearby dwarf galaxy KDG~162 (also known as KK~153) has recently attracted attention as a potentially gas-rich, isolated, and low-mass dwarf galaxy. However, many of these characterizations depend on the true distance to KDG~162. Here we present new Hubble Space Telescope (HST) imaging that resolves individual stars down to 3 magnitudes below the tip of the red giant branch. Using this HST imaging, we determine a distance to KDG~162 of D = 3.48 $\pm$ 0.12~Mpc, somewhat higher than previous determinations. We find KDG~162 to be relatively isolated and lying at the edge of first infall onto the M81 group. We use its stellar content to measure its recent star formation history and find that only a modest amount of star formation can be attributed to within the last $\sim$500~Myr. The present-day stellar mass outweighs the \hi\ mass by a factor of 2$\times$. The properties of KDG~162 add to an increasingly detailed picture of nearby and isolated low-mass galaxies. In the future, deeper observations would allow for reaching more age-sensitive populations, enabling the recovery of its star-formation history back to ancient times.

\end{abstract}


\section{Introduction}

Nearby dwarf galaxies have proven to be extremely useful laboratories for understanding the physics that govern our universe \citep{1999ApJ...522...82K,2017ARA&A..55..343B}. They are generally found to be dark matter dominated, with higher mass-to-light ratios than $L^{*}$ galaxies, and thus useful for investigating the underlying nature of dark matter \citep{2018RPPh...81e6901S,2019ARA&A..57..375S}. Their lower metallicities also make them useful as probes of galaxy formation and evolution at earlier cosmic times \citep{1999AJ....118..302A, 2012A&A...537A.126P, 2023MNRAS.518.5893K}, where individual stars cannot be resolved, and studies must generally be limited to integrated light. 

Furthermore, nearby isolated dwarfs provide testbeds for galaxy evolution that are uncomplicated by factors such as galaxy-galaxy interactions. One of the most prominent examples of such a galaxy is Leo~P, a low-mass, star-forming galaxy located at a distance of D = 1.62 $\pm$ 0.15~Mpc \citep{2013AJ....145..149R, 2015ApJ...812..158M}, placing it at the edge of the Local Group and 0.4~Mpc away from its nearest neighbor (Sextans B). Multi-observatory observations of Leo~P (including ALFALFA, HST, Keck, and the VLA) have revealed Leo P to be an extremely metal-poor, gas-rich, and relatively isolated dwarf galaxy \citep{2013AJ....146...15G,2014AJ....148...35B,2015ApJ...815L..17M,2023ApJ...943...65T}. Deep follow-up observations of Leo~P with JWST/NIRCam have shed light on its ancient star-formation history, indicating a pause in star formation after reionization, with important implications for the UV luminosity function at higher redshifts \citep{2024ApJ...976...60M}. JWST/MIRI observations revealed the presence of cold molecular hydrogen gas at a metallicity of just $3\%$~solar \citep{2025Natur.642..900T}.

In prior decades, searches for nearby dwarf galaxies were primarily conducted by inspecting photographic plates \citep{1990A&A...228...42B, 1998A&AS..127..409K,1999A&AS..135..221K}. These searches progressed to using newer generations of digital imaging \citep{2009AJ....137.3009C,2014ApJ...787L..37M,2014ApJ...795L..35C,2017A&A...597A...7M} as well as \hi\ surveys \citep{2015ApJ...806...95S,2015ApJ...798L..21T} or even backyard amateur telescopes \citep{2016A&A...588A..89J, 2020Ap.....63....5K}. Lately, the business of finding nearby dwarfs continues to be undertaken via similar ongoing investigations as well as by taking advantage of advances in computational techniques \citep{2017ApJ...850..109B,2018ApJ...857..104G,2020ApJ...891..144C,2021ApJ...918...88M,2023Ap.....66..441K,2024ApJ...971L..37J}. 

Among these discoveries, several dwarf galaxies have been found and characterized as likely isolated. These include objects such as UGC~4879 \citep{2011AJ....141..106J}, KKs~3 \citep{2015MNRAS.447L..85K}, Pavo \citep{2023ApJ...957L...5J}, ESO~006-001 \citep{2023ApJ...943..139M}, and Hedgehog \citep{2024ApJ...975L..23L}. Yet another potentially nearby and isolated dwarf galaxy that has attracted recent attention is KDG~162. It was discovered by \cite{1968CoBAO..39...61K} during a search of Palomar Sky Survey prints, and more formally cataloged by \cite{1998A&AS..127..409K} with the name KK~153 (alternatively as LEDA/PGC 41920; \citealt{2003A&A...412...45P}). Further analyses were generally limited to the inclusion of the dwarf in catalogs of potentially nearby galaxies \citep{2000A&AS..141..469H, 2023A&A...678A..16K}. 

More recently, \cite{2025MNRAS.537L..21K} obtained optical spectra for 40 nearby galaxies and measured a radial velocity for KDG~162 of $V_{h}$ = 105 $\pm$ 50 km/s from absorption lines. \cite{2025AJ....170...23N} present Green Bank Telescope (GBT) \hi\ observations of 105 nearby dwarfs including KDG~162, and measured a radial velocity of $V_{h}$ = 125.3 $\pm$ 0.8 km/s, a linewidth of $W_{50}$ = 21.8 $\pm$ 1.9 km/s, and an \hi\ flux of F(\hi) = 0.53 $\pm$ 0.03 Jy km/s. They also performed surface photometry of the dwarf using the DESI Legacy Surveys \citep{2019AJ....157..168D} combined with color transformations to find total apparent magnitudes of B = 17.28~mag and V = 16.65~mag. Assuming a numerical action distance estimate of D = 2.9~Mpc \citep{2017ApJ...850..207S, 2020AJ....159...67K}, they further derived \hi\ and stellar masses of $M_{HI}$ = $10^{6.03}$ $M_{\odot}$ and $M_{*}$ = $10^{6.21}$ $M_{\odot}$, respectively.

KDG~162 was also the subject of \hi\ observations with the Five-hundred-meter Aperture Spherical Telescope (FAST) in \cite{2025ApJ...982L..36X}.  They identified KK~153 as an ultra-faint star-forming dwarf galaxy with a rotating gas disk, and measured $V_{h}$ = 127 $\pm$ 1 km/s, $W_{50}$ = 18 $\pm$ 1 km/s, and F(\hi) = 0.52 $\pm$ 0.05 Jy km/s, all of which closely match the properties given by \cite{2025AJ....170...23N}. Using the baryonic Tully-Fisher relation, they provide a separate estimate of its distance of D =  $2.0^{+1.7}_{-0.8}$~Mpc, indicating it may be a very nearby galaxy (allowing for in-depth follow-up observations), while likely being isolated from any nearest neighbor. Of particular note is that they suggest KDG~162 may be a gas-rich and very low-mass galaxy similar in nature to Leo~P, although these classifications depend on its distance and surface photometry, both of which were brought into question by \cite{2026AA...705A.253B}.

\cite{2026AA...705A.253B} presented Large Binocular Telescope (LBT) observations of KDG~162 with the Sloan Digital Sky Survey (SDSS) \textit{g} and \textit{r} filters. Given the galaxy's relative proximity, they can resolve the brightest individual red giant branch stars. They determine a distance of D = $3.06^{+0.17}_{-0.14}$~Mpc using the tip of the red giant branch technique, which is farther than the distance reported by \cite{2025ApJ...982L..36X}. As noted by \cite{2026AA...705A.253B}, their new distance, combined with their improved integrated photometry, implies a stellar mass estimate for KDG~162 of $M_{*}$ = 2.4 $\pm$ 0.2 $\times$ $10^{6}$ $M_{\odot}$, which makes it comfortably more massive than any reasonable definition for an ultra-faint dwarf ($M_{*}$ $\sim$ $10^{5}$ $M_{\odot}$). Their distance also places KDG~162 in relative isolation ($\sim$630~kpc) from its nearest neighbor.

In this paper, we present Hubble Space Telescope (HST) and Special Astrophysical Observatory (SAO) Large Altazimuth Telescope (BTA) observations that allow us to characterize the stellar populations, star formation, and cosmographic environment of KDG~162 in more detail.

\section{HST Data and Processing}

\begin{figure}
\centering
\includegraphics[width=0.99\linewidth]{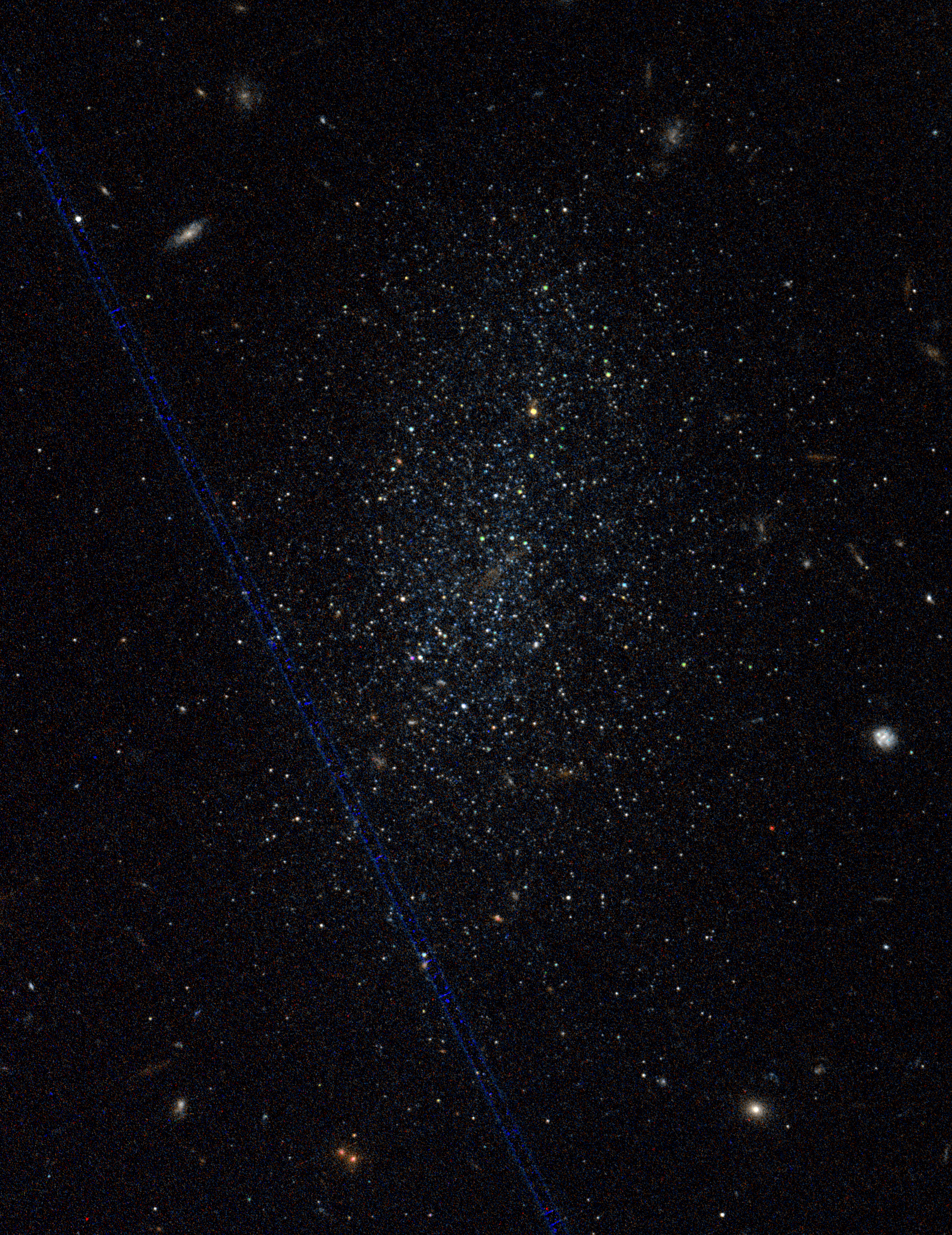}
\caption{HST RGB (F814W/F606W+F814W/F606W) color image cutout (0.9$'\times$1.3$'$) of KDG~162. A modest satellite streak is visible within the underlying F606W data.}
\label{fig:color_image}
\end{figure}

KDG~162 was included in a Cycle 33 HST snapshot survey of suspected Local Volume galaxies \citep{2025hst..prop18070T}, and was observed on December 13 2025 for 748 seconds in each of the F606W and F814W filters with the Advanced Camera for Surveys (ACS). A color image cutout of the galaxy is shown in Figure \ref{fig:color_image}. 

\begin{figure}[!t]
\centering
\includegraphics[width=0.99\linewidth]{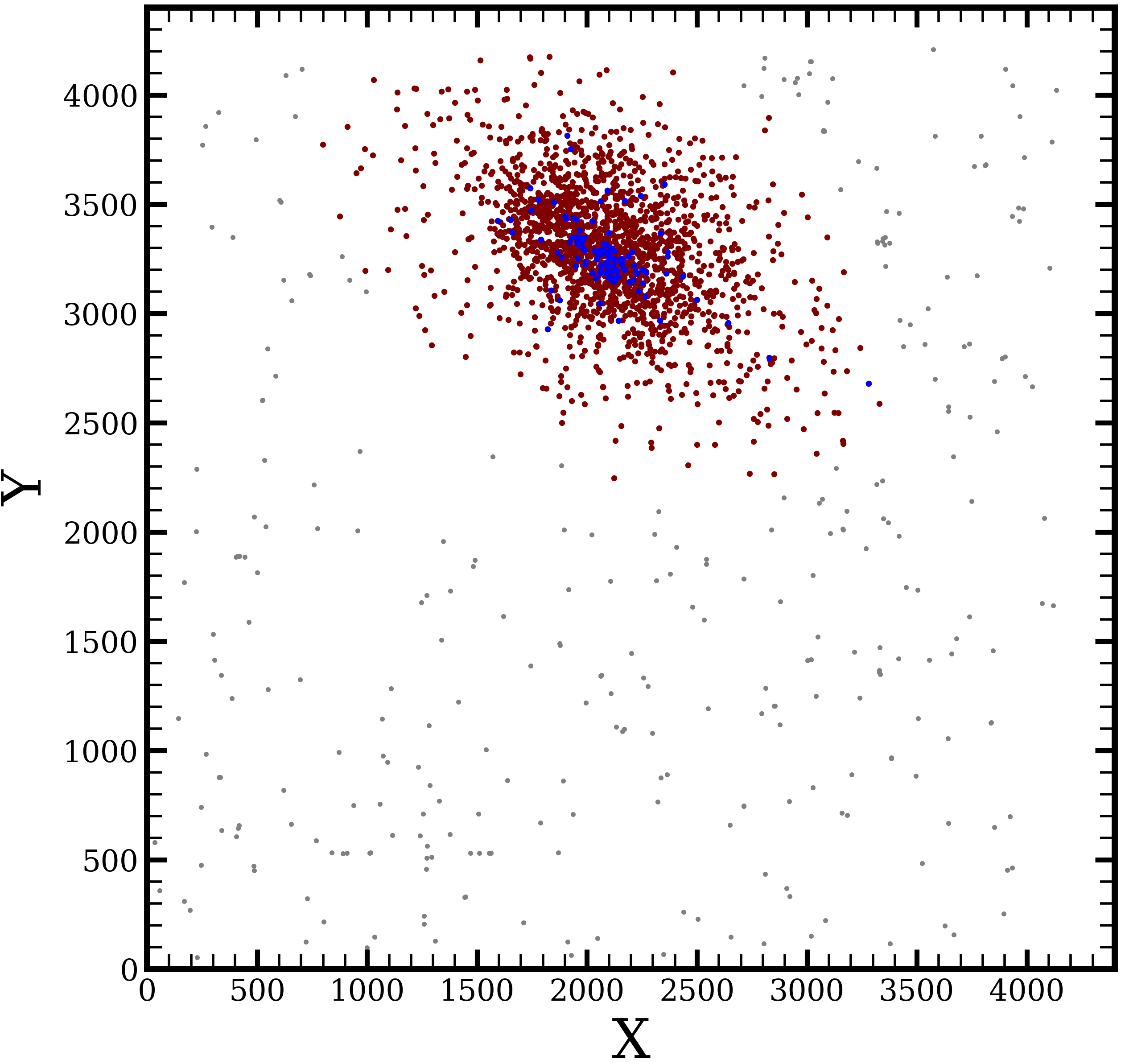}
\caption{The spatial distribution of high-fidelity point sources from our PSF photometry. The sources we associate with the galaxy are shown in red, and those with F606W$-$F814W $<$ 0.1~mag are shown in blue.}
\label{fig:spatial}
\end{figure}

\begin{figure}
\centering
\includegraphics[width=0.99\linewidth]{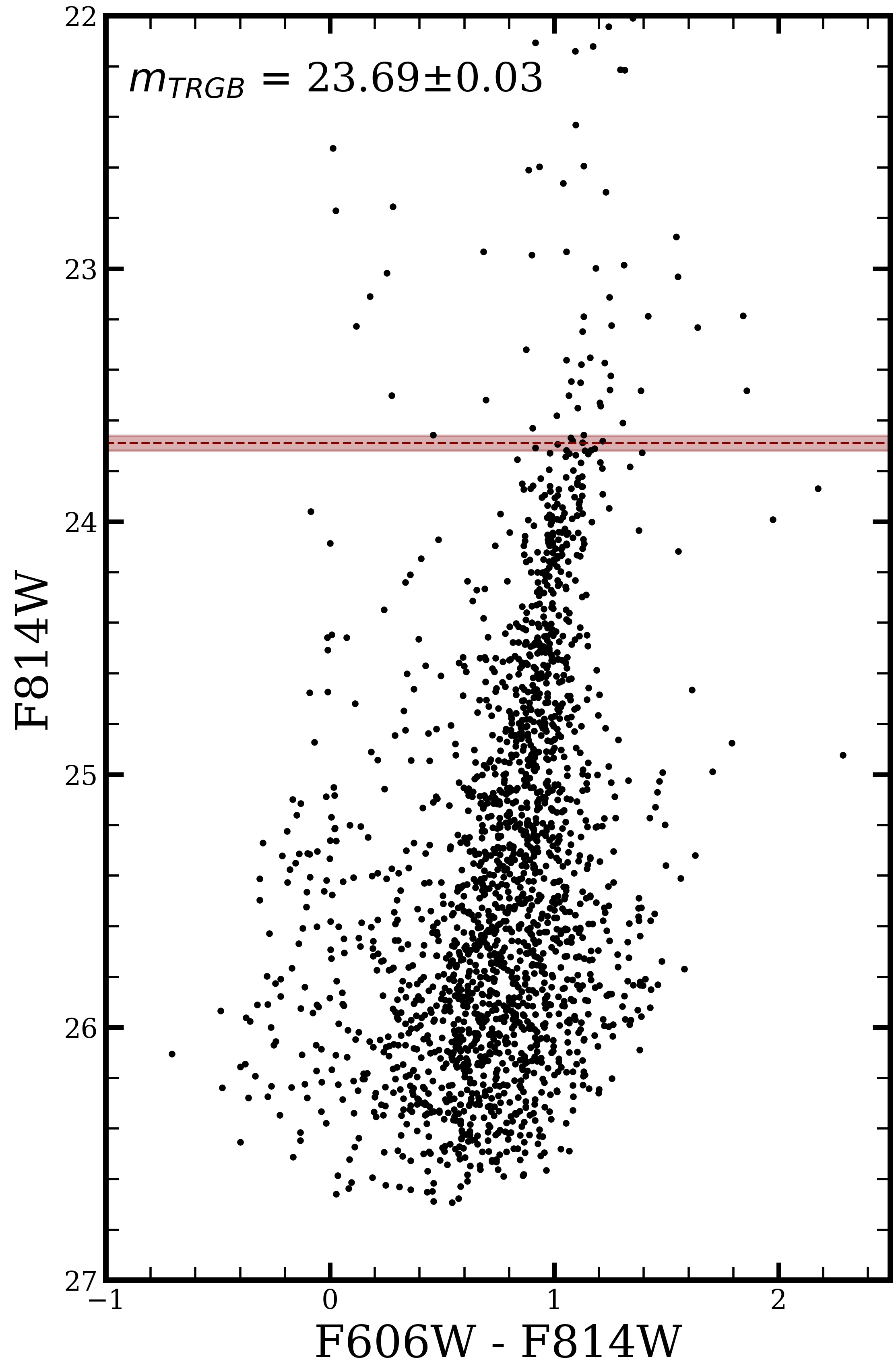}
\caption{Color-magnitude diagram of resolved point sources in KDG~162. The magnitude of the TRGB and its uncertainty is marked in red.} 
\label{fig:cmd_trgb}
\end{figure}

To perform a detailed inventory of the stellar content of the galaxy, we used the DOLPHOT software package \citep{2016ascl.soft08013D} to perform point-spread function (PSF) photometry on the individual \texttt{FLC} exposures, using the combined F814W \texttt{DRC} image as a reference frame for mutual image alignment. We used DOLPHOT reduction parameters recommended by the M31 PHATTER collaboration \citep{2021ApJS..253...53W}, and trimmed the initial output using the following criteria to select only high-fidelity sources \citep{2017AJ....154...51M}: 1) \texttt{Crowd}\textsubscript{F606W} + \texttt{Crowd}\textsubscript{F814W} $<$ 0.8, 2) (\texttt{Sharp}\textsubscript{F606W} + \texttt{Sharp}\textsubscript{F814W})$^{2}$ $<$ 0.075, 3) \texttt{Type} $\le$ 1, 4) S/N $\ge$ 4 (in both filters), and 5) \texttt{Flag} $\le$ 2. The spatial distribution of sources in our field is shown in Figure \ref{fig:spatial}, and the color-magnitude diagram (CMD) of sources that we associate with the galaxy (versus the background) is shown in Figure \ref{fig:cmd_trgb}. The other subset of stars with F606W$-$F814W $<$ 0.1~mag is shown in blue. We also performed artificial star experiments to determine the levels of photometric bias, completeness, and errors. The results from these experiments are provided as input for our distance and star formation history measurements.

KDG~162 exhibits a very low surface brightness and lacks any prominent, 
bright star-forming regions. The upper main sequence (blue stars of about 
F606W-F814W $\lesssim$ 0.1~mag; Fig.~\ref{fig:cmd_trgb}) is only sparsely 
populated and appears to be devoid of sufficiently luminous blue and red 
supergiants, indicating a low level of ongoing star formation activity. Judging 
from the spatial distribution of the blue stars, shown in blue in 
Fig.~\ref{fig:spatial}, the current and recent star formation is compactly 
concentrated toward the central region of the galaxy.

\section{Galaxy Distance}

The HST CMD for KDG~162 shows a prominent red giant population, which we can use to determine its distance via the tip of the red giant branch (TRGB; \citealt{1993ApJ...417..553L, 2021AJ....162...80A}). The method relies on the standard-candle nature of the brightest red giant stars and is based on the physics of the helium flash \citep{Serenelli2017, Beaton2018}. Notably, the subsequent rapid transition to the horizontal branch is marked by a sharp break in the observed CMD. Only modest corrections are required for metallicity in the I-band, hence our use of F814W imaging as the primary filter for the TRGB measurement. To measure the TRGB feature, we follow the methods described in \cite{2006AJ....132.2729M} and \cite{2014AJ....148....7W}. In brief, we isolate the giant branch using a simple color cut (0.5 $<$ F606W-F814W $<$ 1.6~mag) and fit a model luminosity function that takes the form of a broken power law, with the break magnitude corresponding to the TRGB. The results of the artificial star experiments are incorporated to account for photometric bias, errors, and incompleteness. However, we note that these effects are negligible given the relative brightness of the TRGB in our photometry ($\sim$3~magnitudes above the photometric cutoff). We measure a value of $m_{TRGB}$ = 23.69 $\pm$ 0.03~mag. Using the absolute HST TRGB calibration of \cite{2007ApJ...661..815R} shifted 0.05~mag closer to align with the Cosmicflows-4 distance scale \citep{Tully2023}, and accounting for the minor foreground extinction of E(B-V) = 0.008~mag \citep{2011ApJ...737..103S}, we arrive at $\mu$ = 27.71 $\pm$ 0.08~mag, or D = 3.48 $\pm$ 0.12~Mpc.

\section{Star Formation}

\subsection{H$\alpha$ Imaging}

\begin{figure}
\centering
\includegraphics[width=0.99\linewidth]{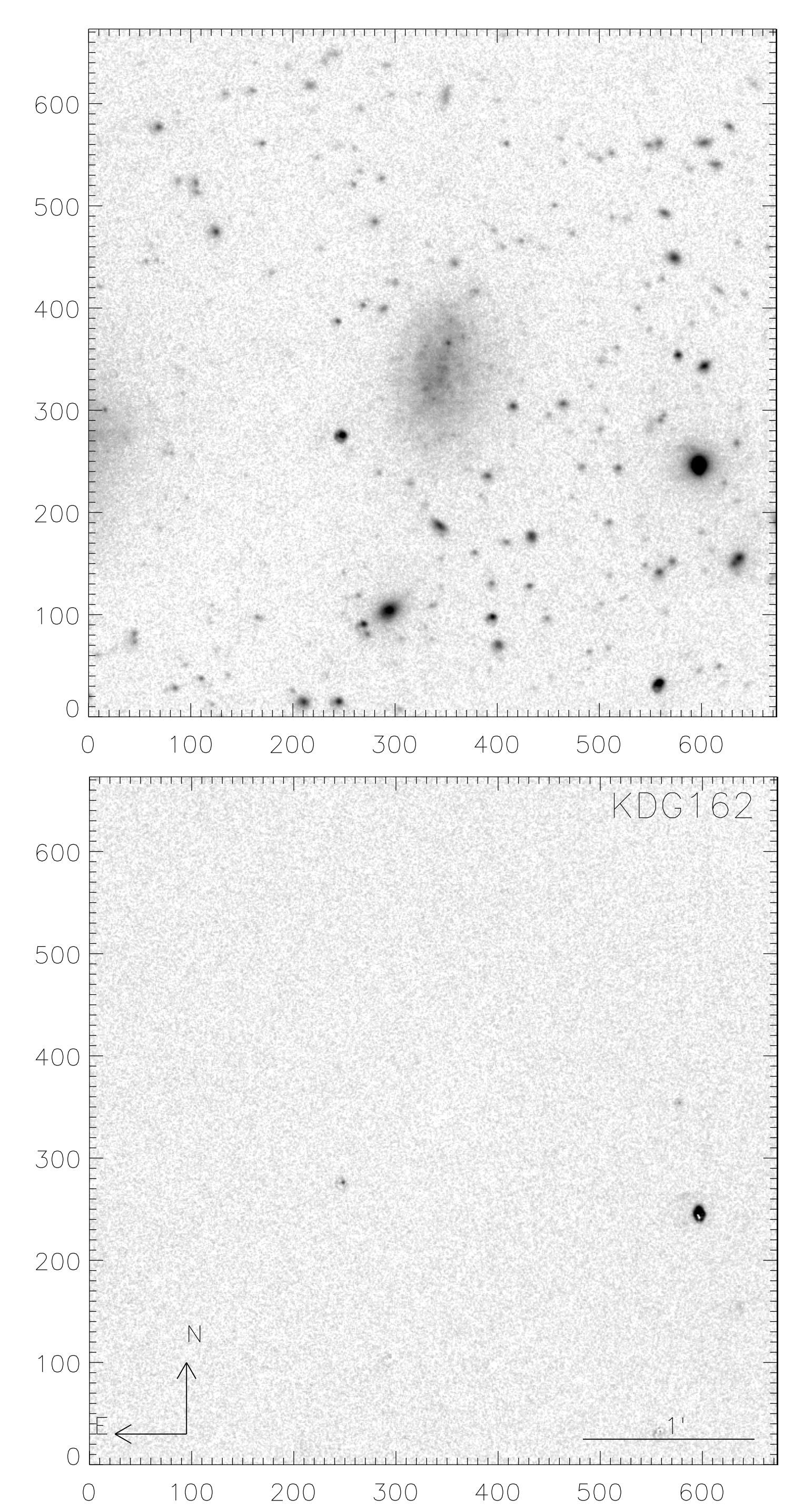}
\caption{\textit{Top}: Medium-band continuum imaging of KDG~162 taken with the Special Astrophysical Observatory's 6-meter BTA \textit{Bottom}: Continuum-subtracted H$\alpha$ imaging of KDG~162, revealing minimal emission. }
\label{fig:halpha}
\end{figure}

KDG~162 was observed with the 6-meter BTA telescope at the Special Astrophysical Observatory in Russia to probe its H$\alpha$ emission. Observations were carried out using the SCORPIO-1 instrument \citep{2005AstL...31..194A} in direct imaging mode on January 30th, 2025. The top panel of Figure \ref{fig:halpha} shows the sum of two medium-band filters, namely \textit{SED608} with $\lambda$=6065$\mathring{A}$ and FWHM=165$\mathring{A}$ (300 seconds) and \textit{SED710} with $\lambda$=7085$\mathring{A}$ and FWHM=204$\mathring{A}$ (300 seconds). The bottom panel shows continuum-subtracted H$\alpha$ emission ($\lambda$=6561$\mathring{A}$ with $\Delta\lambda$=73$\mathring{A}$ for 1200 seconds). We measured a total H$\alpha$ flux of F = 4.5 $\pm$ 2.4$\times$$10^{-16}$ ergs $\mathrm{cm^{-2}}$ $\mathrm{sec^{-1}}$. Using our measured distance to KDG~162, we are able to determine a measure of its current star formation rate (SFR) \citep{1998ARA&A..36..189K,2009ApJ...706..599L,2021MNRAS.506.1346K} with 
\begin{equation}
    \mathrm{log_{10}[SFR(H\alpha)] = 8.98 + 2log_{10}(D) + log_{10}[F(H\alpha)]}.
\end{equation}
We thus arrive at a present-day SFR = $10^{-5.28}$ $M_{\odot}$/yr for KDG~162 as probed via its (minimal) H$\alpha$ emission.

\subsection{Resolved Star Formation History}

\begin{figure}
    \centering
    \includegraphics[width=0.99\linewidth]{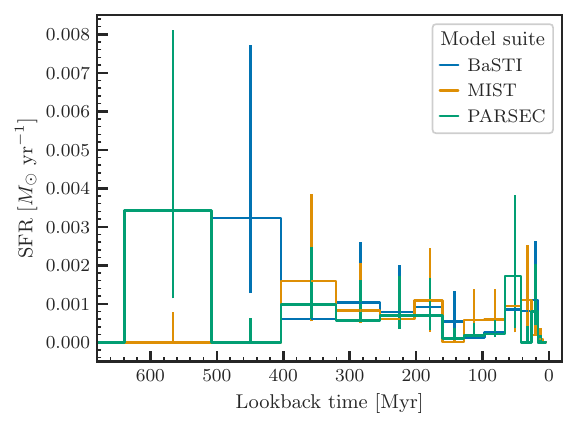}
    \caption{The differential SFH over the past $\sim600$~Myr in solar masses per year, measured with the BaSTI (blue), MIST (orange), and PARSEC (green) stellar models. Steps show the SFR within each time bin, with vertical lines indicating errors.}
    \label{fig:sfh}
\end{figure}

In addition to H$\alpha$, which is sensitive to recent star formation on 10~Myr timescales \citep{1999ApJS..123....3L}, we can estimate the star formation history (SFH) of KDG~162 from its color-magnitude diagram.
For this we use \texttt{MATCH} \citep{2002MNRAS.332...91D}, a well-established software package frequently used for CMD-based SFH fitting \citep[e.g.,][]{
2003ApJ...596..253S, 2009AJ....137..419W, 2010ApJ...721..297M, 
2010ApJ...722.1864M, 2011ApJ...739....5W, 2014ApJ...789..147W, 2015ApJ...805..183L, 2015ApJ...812..158M, 2017ApJ...837..102S, 
2023ApJ...956...86S, 2025ApJ...979..205S, 2025ApJ...992..106D}.
\texttt{MATCH} computes SFH solutions using a Poisson likelihood to compare observed and modeled Hess diagrams.

We model the SFH with three different general-purpose stellar model suites: BaSTI \citep{2004ApJ...612..168P, 2018ApJ...856..125H}, MIST \citep{2016ApJS..222....8D}, and PARSEC \citep{2012MNRAS.427..127B, 2022A&A...665A.126N}.
We fit each model using identical input parameters.
These include a binary fraction of 0.35, an overall metallicity range of $-2.2 < [M/H] < 0.0$ dex with solar-scaled abundances, and an age-metallicity relation constrained to increase with time via the \texttt{-zinc} flag in \texttt{MATCH}. Given the limited depth of the SNAP observations, our CMD does not reach the much more age-sensitive horizontal branch. Hence, we focus our efforts on recovering the recent star formation history. We use an age bin spacing of 0.1 dex between 100~Myr and $\sim$600~Myr, and 0.2~dex below 100~Myr. We do not include ages younger than 16~Myr, as this is the youngest age available in the BaSTI models. 

Although the results for each model fit differ slightly in numerical details, they tell similar stories overall. As can be seen in Figure \ref{fig:sfh}, there is evidence of a recent star-forming episode igniting at least $\sim$400~Myr ago. The average star formation rate was approximately $3 \times 10^{-3} M_\odot\,yr^{-1}$ between 600 and 400 Myr ago, declining to roughly half this value at more recent epochs. The differential SFH exhibits several apparent enhancements and depressions in the star formation rate; however, the associated  uncertainties are comparable to, or even exceed, the measured values.  Therefore, these short-timescale fluctuations should be interpreted with caution and may not represent statistically significant variations. Overall, the reconstructed SFH is consistent with sustained but weak star formation during the last $\sim$500~Myr. Combined with the paucity of luminous  young stars in the CMD, the results may indicate that the present-day star-forming episode is gradually declining, although the uncertainties in the recent SFH prevent a firm conclusion.

\section{Discussion and Summary}

\subsection{Distance}

Our TRGB distance can be compared with the two distances reported in the recent literature in \cite{2025ApJ...982L..36X} and \cite{2026AA...705A.253B}. \cite{2025ApJ...982L..36X} determined a distance of D=$2.0^{+1.7}_{-0.8}$~Mpc using the Baryonic Tully-Fisher relation \citep{2000ApJ...533L..99M, 2022MNRAS.511.6160K}, which, given the relatively high uncertainty of that value, is consistent with our own determination via the TRGB. \cite{2026AA...705A.253B} also measured the distance to KDG~162 using the TRGB, based on observations taken with the Large Binocular Telescope, and found D=$3.06^{+0.17}_{-0.14}$~Mpc. Although the latter distance is closer to our result than that of \cite{2025ApJ...982L..36X}, its smaller uncertainty ($\sim$10$\times$) implies a higher formal discrepancy. The CMD and RGB for KDG~162 in Figure 6 of \cite{2026AA...705A.253B} are somewhat sparse and have a higher potential for contamination from unresolved background galaxies, given the lower resolution of the ground-based imaging. Given that the high-resolution and relatively stable PSF afforded by HST has long been considered the ``gold-standard" for these measurements \citep{2009AJ....138..332J,2017AJ....154...51M,2021AJ....162...80A}, we consider our results to be more robust against potential systematic concerns that may arise from the ground, such as crowding or contamination from the population of background galaxies.

\begin{deluxetable}{l|c|c}[t]
\tablehead{\colhead{\textbf{Parameter}} & \colhead{\textbf{Value}} & \colhead{\textbf{Ref.}}}
\startdata
RA & 12:35:01.6 (J2000) & (1) \\
Dec. & +58:23:08 (J2000) & (1) \\
$A_{V}$ & 0.028 mag & (2) \\
$V_{h}$ & 125.3 $\pm$ 0.8 km/s & (3) \\
$W_{50}$ & 21.8 $\pm$ 1.9 km/s & (3) \\
$\mu$ & 27.71 $\pm$ 0.08 mag & (4) \\
$D$ & 3.48 $\pm$ 0.12 Mpc & (4) \\
$r_{h}$ & 294 $\pm$ 17~pc & (4,5) \\
$M_{V}$ & -11.15 $\pm$ 0.18 mag & (4,5)  \\
$M_{*}$ & 3.1 $\pm$ 0.3 $\times10^{6}M_{\odot}$ & (4,5) \\
$M_{HI}$ & $1.51\pm0.14\times10^{6}$$M_{\odot}$ & (3,4) \\
\enddata
\caption{Summary of key galaxy parameters for KDG~162. Numerical references listed in the table refer to 1) \cite{2013AJ....145..101K}, 2) \cite{2011ApJ...737..103S}, 3) \cite{2025AJ....170...23N}, 4) this work, and 5) \cite{2026AA...705A.253B}.}
\label{tb:parameters}
\end{deluxetable}

\subsection{Stellar and \hi~Mass Estimates}

Our new TRGB distance, combined with previous literature measurements of KDG~162's observed properties, can be used to provide updated galaxy parameters. These are provided in Table \ref{tb:parameters}. The half-light radius and integrated magnitude ($M_{V}$) for KDG~162 are adopted from \cite{2026AA...705A.253B} and rescaled using our new distance. The total \hi\ mass ($1.51\pm0.14\times10^{6}$$M_{\odot}$) is determined using the following relation \citep{1994ARA&A..32..115R}:
\begin{equation}
    M_{HI} = 2.36 \times 10^{5} D^{2} \int S dV
\end{equation}
with our distance (D) and the \hi\ flux ($\int S dv$) measured by \cite{2025AJ....170...23N}. 

\cite{2026AA...705A.253B} also provides an estimate of $M_{*}$ using their integrated light measurements and optical (g and r-band) mass-to-light ratios \citep{2015MNRAS.452.3209R,2016AJ....152..177H}. They determine a value of $M_{*}$ = 2.4 $\pm$ 0.2 $\times$$10^{6}$ $M_{\odot}$, although they also note that uncertainties tied to the underlying mass-to-light ratios are at least as large as $\pm$20$\%$. Simply scaling their measurement to our new distance yields an updated value of $M_{*}$ = 3.1 $\pm$ 0.3 $\times$$10^{6}$ $M_{\odot}$. Using this updated value, we determine $M_{HI}/M_{M*}=$ 0.5. Given our results, it appears clear that KDG~162 is not gas-rich ($M_{HI}/M_{M*}>1$). \cite{2025ApJ...982L..36X} referred to KDG~162 as both gas-rich and ultra-faint ($M_{*} \lesssim 10^{5}$ $M_{\odot}$), which \cite{2026AA...705A.253B} determined was not the case based on their updated distance and improved integrated-light measurements. Our measurements reinforce the case that KDG~162 meets neither of these criteria.

\subsection{Cosmography}

\begin{figure*}
\centering
\includegraphics[width=0.99\linewidth]{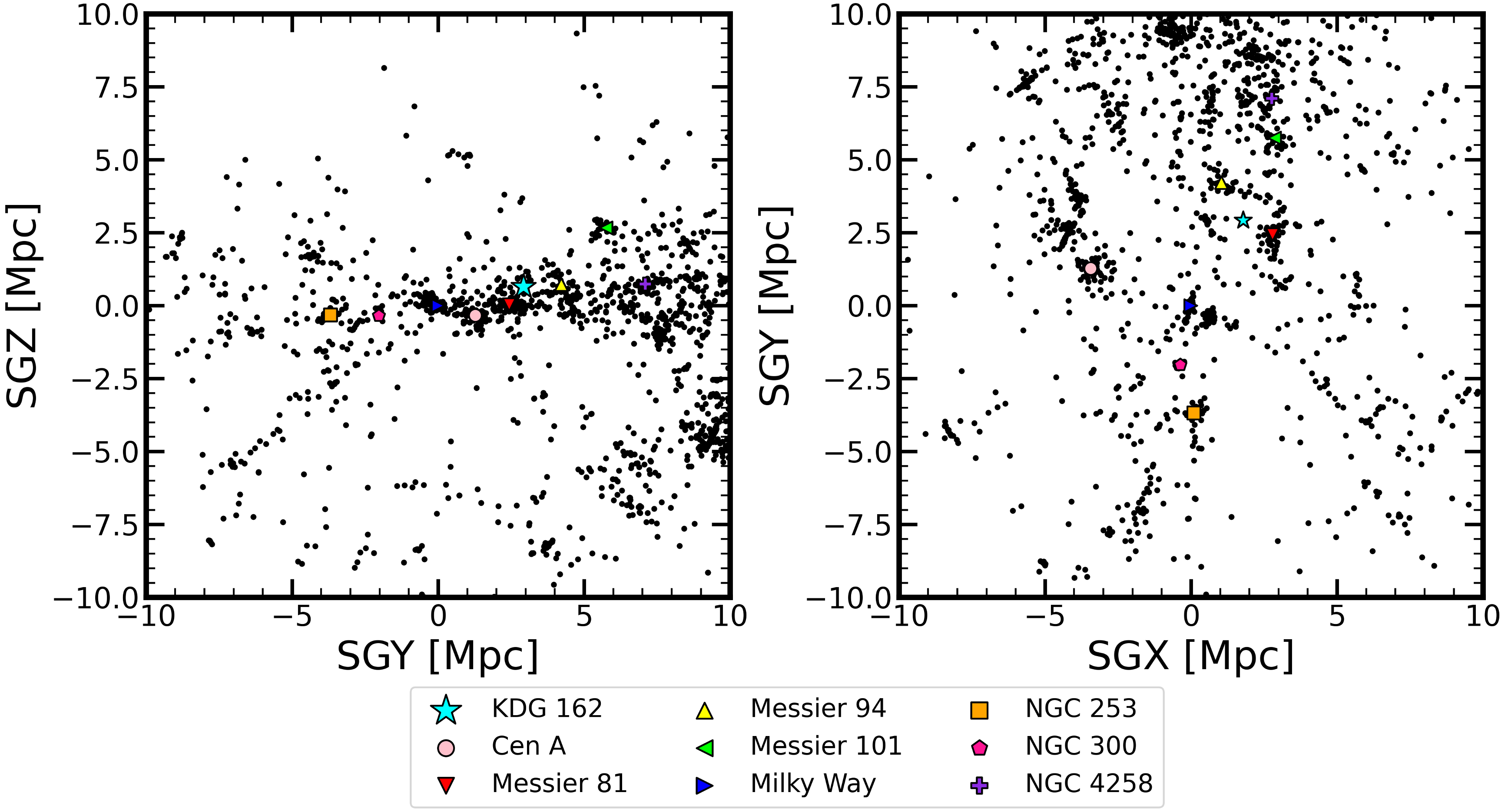}
\caption{\textit{Left:} Plot of galaxies in the online version of the Updated Nearby Galaxy Catalog, displayed in the supergalactic coordinate system. KDG~162 is shown as the cyan star. Several other key galaxies in the Local Volume are also highlighted for context. \textit{Right:} Same as the left-hand side, but now in the SGX-SGY plane. The relative isolation of KDG~162 can be more easily seen in this view.}
\label{fig:sg}
\end{figure*}

\begin{figure}
\centering
\includegraphics[width=0.99\linewidth]{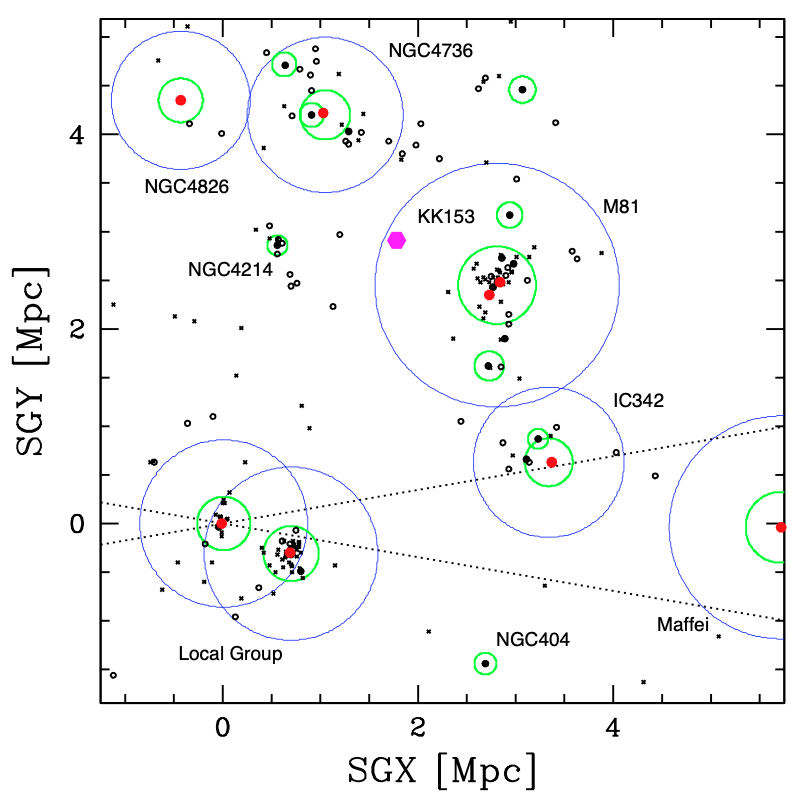}
\caption{Zoom-in of the region around KDG~162/KK~153 in the supergalactic SGX-SGY plane. Galaxies in red represent the most massive nearby galaxies. The green circles represent splashback radii, whereas the blue circles denote infall radii. KDG~162 lies just within M81's infall region.}
\label{fig:sg_zoom}
\end{figure}

Our highly precise distance to KDG~162 allows us to examine its cosmographic environment in detail. Figure \ref{fig:sg} shows two views of the nearby universe in the supergalactic coordinate system \citep{1956VA......2.1584D}. The galaxies, positions, and their distances are obtained from the online and regularly updated version\footnote{\url{https://www.sao.ru/lv/lvgdb/introduction.php}} of the Updated Nearby Galaxy Catalog \citep{2013AJ....145..101K}, known as the Catalog of the Local Volume (LV) Galaxies. The left-hand panel shows the SGY-SGZ plane, and the right-hand panel the SGX-SGY plane. Most nearby galaxies lie within a thin plane at SGZ$\sim$0, also known as the Local Sheet. Few galaxies are located at positive SGZ, which is the domain of the Local Void \citep{2019ApJ...880...24T}. Using our newly derived distance, KDG~162 is found to lie within the bounds of the Local Sheet. The SGX-SGY view affords a view in the plane of the Local Sheet, and helps showcase the relative isolation of KDG~162. KDG~162 is separated from its nearest neighbor UGC~7298 by a distance of 830~kpc. 

Figure \ref{fig:sg_zoom} provides a more restricted view of the SGX-SGY plane using distances from the CMDs/TRGB Catalog \citep{2009AJ....138..332J, 2021AJ....162...80A}. Massive galaxies are shown as red points, with their splashback ($r_{2t}$) and infall ($r_{1t}$) radii \citep{2015AJ....149...54T} denoted with green and blue circles, respectively. It can be seen that KDG~162 is within the infall region surrounding M81, indicating that it has not yet been perturbed by the halo of M81. Its local environment indicates that its evolution can be examined in the context of a relatively isolated galaxy.

\subsection{Star Formation History in Context}

\cite{2024ApJ...976...60M} present a deep CMD-based SFH for Leo P, which, as mentioned earlier, is a gas-rich, low-mass, and isolated system. They compare the SFH of this system to three other nearby galaxies, which also have a CMD-based SFH (reaching down to the oMSTO) and meet the same additional criteria. These systems are Aquarius \citep{2014ApJ...795...54C}, Leo A \citep{2007ApJ...659L..17C}, and WLM \citep{2024ApJ...961...16M}. In this comparison, they find a common theme of a pause in star formation post-reionization, followed by its re-ignition at later times. 

\cite{2026ApJ...997..271M} provide three additional points of comparison by including the recently discovered galaxies Pavo \citep{2023ApJ...957L...5J}, Corvus~A \citep{2024ApJ...971L..37J}, and Kamino \citep{2026ApJ...997..271M}. Pavo and Kamino do not conform to the same trend as those in \cite{2024ApJ...976...60M}, and instead show a more continuous pattern of star formation without strong evidence of a post-reionization pause. However, this comparison is complicated by the fact that the imaging for these newly discovered galaxies is significantly shallower and does not meet the criteria of reaching the oMSTO. Additionally, Pavo does not meet the same definition of being gas-rich, with an \hi\ -to-stellar mass ratio of $M_{HI}/M_{M*}=0.5$, while the same quantity for Kamino is uncertain due to a lack of targeted \hi\ data. 

Other nearby systems with lower \hi\ -to-stellar mass ratios than the gas-rich systems analyzed within \cite{2024ApJ...976...60M} include UGC~4879 \citep{2009AJ....138..332J} and ESO~006-G001 \citep{2023ApJ...943..139M}. In these cases, it appears that the bulk of their stellar mass had formed early on in the universe's lifetime (lookback times, or $t_{L}$$\sim$10$-$14~Gyr), followed by very limited intermediate-time ($t_{L}$$\sim$4$-$10~Gyr) star formation, and a modest amount of star formation at the most recent epochs ($t_{L}$$\sim$0$-$4~Gyr). Such behavior is reminiscent of nearby dwarf transitional (dTr) galaxies. \citet{2021MNRAS.502.1623M} considered the residual SF in nearby dTr/dSph galaxies depending on the degree of isolation of the object. Highly isolated dwarfs clearly show recent residual SF ($\ge$10 percent of the total stellar mass), while for those within the virial radius of massive neighbors, it is practically zero. 

\subsection{Summary}

The nearby dwarf galaxy KDG~162 \citep{1968CoBAO..39...61K}, also known as KK~153 \citep{1998A&AS..127..409K}, had begun to attract more recent attention due to claims that it may be an isolated, very low-mass, and gas-rich galaxy \citep{2025ApJ...982L..36X} similar to Leo P \citep{2013AJ....146...15G, 2015ApJ...812..158M}. KDG~162 was recently observed as part of an ongoing HST snapshot program of suspected Local Volume galaxies. We use this HST imaging to confirm many of the recent findings of \cite{2026AA...705A.253B}, albeit with some differences in the details. We find that the distance to KDG~162 is farther than that measured by both \cite{2025ApJ...982L..36X} and \cite{2026AA...705A.253B}, implying that KDG~162 is significantly more massive than Leo~P. 

Combined with improved integrated light measurements \citep{2026AA...705A.253B} and stellar mass estimates, we also find that KDG~162 does not meet the criteria for being gas-rich ($M_{HI}/M_{M*} >1$), with $M_{HI}/M_{M*}=$ 0.5. Our new distance places KDG~162 significantly outside the Local Group, and just within the bounds for an eventual infall onto the M81 group. We also use our CMD to derive the recent SFH of KDG~162 and find a modest amount of star formation within the last $\sim$600~Myr. Deeper observations of a larger, more varied sample of isolated, low-mass dwarf galaxies including KDG~162 would help reconstruct their SFHs back to ancient times and provide valuable insight into the apparent variations in their evolutionary pathways.

\begin{acknowledgments}
G.S.A acknowledges support from HST GO-18070. I.D.K. and L.N.M. are supported by the Russian Science Foundation grant No 24-12-00277, ``Cosmography of the Local Volume of the Universe within 15 megaparsecs".
\end{acknowledgments}

\facilities{HST/ACS, SAO/BTA 6m}

\software{\texttt{DOLPHOT} \citep{2016ascl.soft08013D}, \texttt{MATCH} v2.7 \citep{2002MNRAS.332...91D}}

\bibliography{main,meredith}
\bibliographystyle{aasjournalv7}

\end{document}